\documentstyle[12pt]{article}

 \def\+{{+\!\!\!+}}

 \def\d{\partial}

 \def\pmb#1{\setbox0=\hbox{#1}%

\kern.0em\copy0\kern-\wd\kern-.04em\copy0\kern-\w\kern.08em\copy0\kern
 -\wd0

 \kern-.04em\raise.0433em\box0 }         
 
 \newcommand{\nc}{\newcommand}
 \nc{\beq}{\begin{equation}}
 \nc{\eeq}[1]{\label{#1}\end{equation}}
 \nc{\ber}{\begin{eqnarray}}
 \nc{\eer}[1]{\label{#1}\end{eqnarray}}
 \nc{\pek}[1]{\cite{#1}}
 \nc{\enr}[1]{(\ref{#1})}
 \nc{\kal}[1]{{\cal{#1}}}
 \nc{\dott}{\;\cdot\;}

\begin{document}
 \newcommand{\inv}[1]{{#1}^{-1}} 
 \renewcommand{\theequation}{\thesection.\arabic{equation}}
 \newcommand{\be}{\begin{equation}}
 \newcommand{\ee}{\end{equation}}
 \newcommand{\bea}{\begin{eqnarray}}
 \newcommand{\eea}{\end{eqnarray}}
 \newcommand{\re}[1]{(\ref{#1})}
 \newcommand{\qv}{\quad ,}
 \newcommand{\qp}{\quad .}
 \begin{center}
                                  \hfill UUITP-08/03\\
                                  \hfill   hep-th/0305098\\
 \vskip .3in \noindent

 \vskip .1in

 {\large \bf{Tensionless Strings, WZW Models at
 Critical Level \\
 and \\
   Massless Higher Spin Fields}}
 \vskip .2in

 {\bf Ulf Lindstr\"om}$^a$\footnote{e-mail address:
 ulf.lindstrom@teorfys.uu.se}
 and  {\bf Maxim Zabzine}$^{b}$\footnote{e-mail address:
 zabzine@fi.infn.it; Address after September 1, 2003: LPTHE, Universit\'es Paris VI et VII,
 4pl Jussieu, 75252 Paris, Cedex 05 France}
 \\


 \vskip .15in

 \vskip .15in
 $^a${\em Department of Theoretical Physics, \\
 Uppsala University,
 Box 803, SE-751 08 Uppsala, Sweden }\\
 \vskip .15in
 $^b${\em INFN Sezione di Firenze, Dipartimento di Fisica\\
 Via Sansone 1, I-50019 Sesto F.no (FI), Italy}

 \bigskip


 \vskip .1in
 \end{center}
 \vskip .4in

 \begin{center} {\bf ABSTRACT }
 \end{center}
 \begin{quotation}\noindent
 We discuss the notion of tensionless limit in quantum bosonic string
 theory, especially
   in flat Minkowski space, noncompact group manifolds (e.g.,
 $SL(2,R)$) and coset   manifolds (e.g., $AdS_d$).
 We show that in curved space typically there exists a critical
 value of the tension which
 is related to the critical value of the level of the corresponding affine
 algebra.
 We argue that at the critical level the string theory becomes
 tensionless and that there exists a
 huge new symmetry of the theory.
 We discuss the appearance of the higher spin massless states at
 the critical level.
 \end{quotation}
 \vfill
 \eject


 \section{Introduction}

 The aim of this paper is to initiate the discussion of the tensionless
 limit (i.e., $\alpha' \rightarrow
 \infty $ or $T = (2\pi \alpha') \rightarrow 0$)  in the quantized
 string theory.
   The tensionless limit of string theory should give
 us an idea about a short-distance properties of the theory. Naively in this
 limit all particles
 will have vanishing mass and therefore new symmetries should
 appear. This has previously been shown to be the case for the {\em classical}
tensionless string \cite{Karlhede:wb}, and its quantized version in \cite{Isberg:1992ia},
\cite{Isberg:1993av},\cite{Gustafsson:1994kr} in a flat background.

However a priori
 it is an open question if this limit gives rise to a consistent theory.
 In this letter we would
 like to argue that there {\em is} such limit for some target spaces and that the
 theory would have new
 symmetries associated with higher spin massless particles.

 Unlike the previous studies of the limit cited above, we would like to
 consider the tensionless
 limit directly in the quantum string theory. Since the nature of the limit is
 highly quantum, in the path integral small tension corresponds to large $\hbar$,
 this is a natural thing to do. We shall be interested in the general case when the target space
is a curved manifold. Our main examples will be group and coset
 manifolds.

 An important incentive for the present study comes from from recent discussions of the $AdS/CFT$
 correspondence
 at vanishing Yang-Mills coupling constant \cite{Sundborg:2000wp}. On the string side
 this correspondence
 requires taking the tensionless limit.

   The paper is organized as follows. In Section \ref{s:flat} we
 consider the tensionless
 limit in flat Minkowski space. We discuss the limit at the level
 of Hilbert space and
 Virasoro constraints. We show that in the tensionless limit the
 Virasoro constrains give
 rise to Fronsdal's conditions for  free massless high spin fields.
 However it is highly probable
 that higher spin massless interactions cannot be constructed in
 flat space and hence that the tensionless limit is inconsistent in
 flat space as an interacting theory.   In Section \ref{s:group}
 we turn to the discussion of string theory on noncompact group
 manifolds. In this case the level $k$
   of the corresponding  affine algebra can be identified with a
 dimensionless analog of the string
 tension ($k=2\pi TR^2$ where $R$ is the size of a group manifold).
For the
 theory to be unitary
 the level is typically restricted to $ - h^V < k < \infty$
 where $h^V$ is the dual
 Coxeter number. We argue that the tensionless limit corresponds to
 taking the level to a
 critical value (i.e., $k = - h^V$) where the number of zero-norm
states
 increases dramatically
 and thus indicate the appearance of new huge gauge symmetry.
 Finally in Section \ref{s:coset} we discuss the tensionless limit
 for coset manifolds,
 in particular, we consider $AdS_d$ space and we show how the free
 massless high spin fields may  arise in the limit.
 A summary of our results and comments regarding the future directions of
 investigation are
 collected in Section \ref{s:end}.

 In this letter we consider the bosonic string and ignore the
 questions of consistency of
 the theory. However we believe that similar results will  hold for
 suprestrings.

 \section{Tensionless strings in flat space}
 \label{s:flat}

 In this section we consider the tensionless limit for the bosonic
 string in flat Minkowski
 space. Despite the fact that the subject has been around for 15
years, we think
 that some points have
 been overlooked. Besides the flat space example serves as a good
 starting point for a
 discussion of string theory on curved manifolds.

 Let start from the standard bosonic string action in conformal gauge
 living in
   flat space
 \beq
   S= \frac{1}{4\pi\alpha'} \int d^2\sigma \, \eta_{\mu\nu} \d_\alpha
    X^\mu \d^\alpha X^\nu ,
 \eeq{actionflat}
 where $\eta_{\mu\nu}=diag(-1,1,...,1)$. The parameter in front of the
 action
   is  the string tension $T=(2\pi\alpha')^{-1}$. In this
section
 we discuss some aspect of string theory when the tension $T$ is
 taken to zero, i.e.
    tensionless strings \cite{Karlhede:wb},\cite{Schild:vq}. This can be done in different ways and it has
 been discussed extensively
 in the literature. For example, one can take the limit at the level
 of classical action
 (\ref{actionflat}) \cite{Lindstrom:1990ar},\cite{Lindstrom:1990qb} and then quantize it \cite{Isberg:1992ia}
\cite{Isberg:1993av},
\cite{Gustafsson:1994kr}.
Another
 approach is
 to consider the limit at the level of scattering amplitudes \cite{Gross:1987kz}-\cite{Amati:1988tn}.

 However here we discuss the tensionless limit in the
 free quantum theory,
   at the level of the Hilbert space. For the present discussion it is
 enough to
    work within the old covariant quantization program (for review,
see
 \cite{Green:sp}).
 For the sake of simplicity we consider the open string. However the
 whole discussion can
 be straightforwardly generalized to closed strings.
   The field $X^\mu$ is expanded in modes which obey the
 commutation relations
 \beq
   [a^\mu_n, a^\nu_m]=\eta^{\mu\nu}\delta_{n+m},\,\,\,\,\,\,\,[q^\mu,
 p_\nu] = i\delta^\mu_{\,\,\nu}
 \eeq{comutrel}
 where $(a^\mu_n)^{\dagger}=a^\mu_{-n}$. The Fock space is built by
 the actions
   of $a_{-n}^\mu$ with $n > 0$ on the vacuum $|0,k\rangle$ such that
 $p_\mu |0,k\rangle
    =k_\mu |0, k\rangle$. Physical states are those that  satisfy the
 Virasoro
     constraints
     \beq
      (L_0 - 1) |phys \rangle =0,\,\,\,\,\,\,\,\,\,\,\,\,\,\,\,
      L_n |phys \rangle =0,\,\,n > 0
     \eeq{VIR}
 where the Hamiltonian $L_0$ is
 \beq
 L_0 = \frac{1}{2} \alpha' p_\mu \eta^{\mu\nu} p_\nu +
 \sum\limits_{n\neq 0}
   n (a^\mu_{n})^\dagger \eta_{\mu\nu} a^\nu_n
 \eeq{L0flat}
 and the $L_n$'s are
 \beq
   L_n = \sqrt{\alpha'} p_\mu a^\mu_n + \sum\limits_{m=1}^{\infty}
 \sqrt{m (m+n)}
    a_{n+m}^\mu \eta_{\mu\nu} (a^\nu_m)^\dagger +\frac{1}{2}
     \sum\limits_{m=1}^{n-1} \sqrt{m (n-m)} a_m^\mu \eta_{\mu\nu}
 a^\nu_{n-m}
 \eeq{Lnflat}
   It is important that our expressions are properly normalized. The
 momentum $p$ requires
 $\sqrt{\alpha'}$ and the creation and annihilation operators are
 taken dimensionless.
 Indeed Gross and Mende have used the same normalization in their
 study of high-energy string
 scattering \cite{Gross:1987kz}, \cite{Gross:1987ar}, \cite{Gross:1988ue}.

   Next we take the tensionless limit (i.e., $\alpha' \rightarrow
 \infty$)  at
    the level of Virasoro constraints. The tensionless limit will
 result in modifications
     of the constraints (\ref{VIR}), namely
     \beq
       (p_\mu\eta^{\mu\nu} p_\nu) |phys \rangle =
 0,\,\,\,\,\,\,\,\,\,\,\,\,
       (p_\mu a_n^\mu) |phys \rangle = 0,\,\, n> 0 .
     \eeq{newconflat}
     These constraints, $l_0 =p_\mu \eta^{\mu\nu} p_\nu$ and $l_n =
 p_\mu a_n^\mu$
      for $n > 0$ together with $l_{-n} = l^\dagger_n$, do not generate
      the Virasoro algebra, instead the corresponding  algebra
 is
      \beq
       [l_n, l_m] = \delta_{n+m} l_0,\,\,\,\,\,[l_n,l_0]=0,\,\,\,\,\,
       n\neq 0,\,\,m\neq 0
      \eeq{HEalg}
      which is the Heisenberg algebra with $l_0$ being the central
element.
 Now we can analyze the string spectrum using the new conditions on
 the physical
 spectrum. Following the standard prescription we construct the Fock
 space using the creation operators,
   the negative norm states are supposed to be projected out by the
new
 conditions (\ref{newconflat})
 and the physical states will be organized according to massless
 representations of the Poincar\'e group.
 Indeed for some of the states these mass-shell and transversality
 conditions (\ref{newconflat})
 give us the Fronsdal's massless free
    higher spin fields (in the specific on-shell gauge)\footnote{Previously the realization of the higher
 spin symmetries in free string field theory has been discussed in \cite{Francia:2002pt}.}.
  To illustrate this point we  consider as an
 example the sector
     build from $a_{-1}^\mu$. The Poincar\'e irreducible
 representation of spin $s$
      corresponds to
   \beq
    |\phi \rangle = \epsilon_{\mu_1 ...\mu_s}(k) a_{-1}^{\mu_1} ...
 a_{-1}^{\mu_s}         |0,k\rangle~,
   \eeq{exhighsp}
    where $\epsilon_{\mu_1 ...\mu_s}(k)$ is a symmetric and traceless
 field (i.e., $\eta^{\mu_1\mu_2}
 \epsilon_{\mu_1\mu_2...\mu_s}= 0$) and therefore the representations
 of
   the corresponding flat space little group  $O(d-2)$ are
 characterized by Young tableaux with one row.
     The conditions (\ref{newconflat}) ensure that we are working with
 free
      massless higher spin fields,
 \beq
 k^{\mu} \eta_{\mu\nu} k^{\nu} = 0,\,\,\,\,\,\,\,\,\,\,\,\
 k^{\mu_1} \epsilon_{\mu_1 ...\mu_s}(k) =0 .
 \eeq{condphi}
 The second condition in (\ref{condphi}) should be interpreted in same
 way as it
 done in QED when by imposing condition $\d_\mu A^\mu = 0$ on the Fock
 space one
 kills unwanted states.
 The gauge transformations amount to a shift of the state by a null
 state (a physical state
 which is orthogonal to all physical states and therefore of zero
 norm):
 \beq
   |\phi \rangle \,\,\,\,\longrightarrow\,\,\,\,|\phi \rangle +
 k_{\mu_1} \gamma_{\mu_2 ...\mu_s} a_{-1}^{\mu_1} ... a_{-1}^{\mu_s}
 |0,k\rangle~.
 \eeq{gaugetr}
 Here $k^{\mu_2} \gamma_{\mu_2 ...\mu_s} = 0$ and
 $\gamma$ is
 completely symmetric tensor by construction. Obviously in (\ref{gaugetr}) the shifted
 state has
 zero norm on shell, where $k^2=0$. Alternatively we may rewrite the
 transformation (\ref{gaugetr})
 as follows
 \beq
   |\phi \rangle \,\,\,\,\longrightarrow\,\,\,\,|\phi \rangle +
 l_{-1} \gamma_{\mu_2 ...\mu_s} a_{-1}^{\mu_2} ... a_{-1}^{\mu_s}
 |0,k\rangle .
 \eeq{gaugetr1}
 In a similar fashion the other states in the Fock space may be analyzed.
 The important new property
 is that the number of null states is huge. For example, all states of
 the form
 \beq
 l_{n_1} l_{n_2} ... l_{n_p} |0,k\rangle,\,\,\,\,\,\,\,n_i >
 0,\,\,i=1,...,p
 \eeq{nullflat}
 are null states.
 Thus we witness the appearance of a new large  symmetry which corresponds to the
 gauge symmetries of
 massless higher spin free fields. A similar conclusions regarding
  the appearance of new symmetries have been
 made by Gross \cite{Gross:1988ue}
 in studying high-energy string scattering.

 In this naive tensionless limit we see that  there are
 massless
   free high spin fields in the spectrum. However we know that there is no consistent
 interacting theory
  for  these fields in flat Minkowski space.\footnote{Interacting higher spin theories
 typically require a non-zero cosmological constant, and have been extensively studied,
starting in \cite{Fradkin:ks}. For recent progress, see  \cite{Vasiliev:2003ev} and references therein.}
 Therefore we
 conclude that the present tensionless limit does not produce
      a consistent (non-free) theory.  We should keep in mind, however,
that in drawing the conclusion that no interacting
theory exists, use is generally made of the
 Coleman-Mandula theorem \cite{Coleman:ad}, which in turn is proven under the assumption of
 a finite number of different particles \cite{Gross:1988ue}.

 \section{Tensionless strings on group manifolds}
 \label{s:group}

 From this section onwards, we discuss the notion of a
 tensionless
   limit in the setting of curved space. We try to repeat the idea from
the
 previous
    section in that we first construct the tensionfull quantum theory and
 then
     only at the quantum level take the tensionless limit.

     To begin with, we consider the tensionless limit of
 strings on
      group manifolds. The main example we have in mind is $SL(2,R)$,
 however
       most of the discussion goes through for other noncompact
groups.

     Let us consider the sigma model (i.e., the gauge fixed string
 action)
   over a group manifold ${\cal G}$ with a Lie algebra ${\bf g}$
     \beq
      S= \frac{1}{4\pi\alpha'} \int d^2\sigma\,\, (g_{\mu\nu} +
 B_{\mu\nu})
       \d_+ X^\mu \d_- X^\nu,
     \eeq{sigmamodel}
      where $X^\mu$ is coordinate on the target space (i.e.,
 dimensionful field).
       With $R$ a dimensionful parameter which characterizes the size
 of the manifold,
        we can rescale $\phi^\mu = R X^\mu$ such that
 $\phi$ is dimensionless. Using this dimensionless
         field $\phi^\mu$ we rewrite the action in terms of group
 elements $g$
         \beq
          S= \frac{k}{4\pi} \int d^2 \sigma \,Tr (g^{-1} \d_\alpha g
 g^{-1}
           \d^\alpha g) + \frac{k}{12\pi} \int d^3\sigma\,
 \epsilon^{\alpha
           \beta\gamma}Tr( g^{-1}\d_\alpha g g^{-1}\d_\beta g
 g^{-1}\d_\gamma g)
         \eeq{WZWmodel}
      where $k= R^2/(\alpha')$ is the level. Thus for a group
 manifold $k$
       is a dimensionless analog of the string tension \cite{Lindstrom:1999tk} and therefore,
 classically the
    tensionless limit amounts to taking the level $k$ to $0$.
 However, we believe that this is not in general an allowed limit at the
 quantum level.

   For compact groups the level $k$ is quantized and should be
 positive.
    Thus we cannot take it continuously to zero, and the smallest
 possible positive
      level in the theory does not have any special properties. Therefore
 we conclude that there are no meaningful tensionless limits for
compact
        group manifold. This should not come as a surprise
 since  massless particles on a compact manifold are problematic.

 However if the group is noncompact then typically the level $k$ is
 not quantized.
 In this case only the positivity of the central charge
 restricts the allowed values of $k$. To understand
these restrictions,
 let us spell out the steps in the Sugawara construction.
  For the WZW model the affine symmetry is given by the
Kac-Moody algebra
 \beq
 [J_n^A , J_m^B ] = i f^{AB}_{\,\,\,\,\,\,\,\,C} J_{n+m}^C + k
 \eta^{AB}n\delta_{n+m}
 \eeq{KMalg}
 where $f^{AB}_{\,\,\,\,\,\,\,\,C}$ are the structure constants of
${\bf g}$.
 We define the Sugawara operators as follows
 \beq
 l_n = \frac{1}{2}\sum\limits_{m = -\infty}^{+\infty} : J^A_m
 \eta_{AB} J^B_{n-m}:
 \eeq{definS}
  where dots denote normal ordering.
 They satisfy the the following commutation relations
 \beq
 [l_n, J_m^A ] = - (k + h^V) m J_{n+m}^A ,
 \eeq{SJJ}
 \beq
 [l_n, l_m] = (k + h^V) \left ( (n-m) l_{n+m} + \frac{k \dim{\bf
g}}{12}(n^3-n)\delta_{n+m}\right ) ~,
 \eeq{SSL}
where $h^V$ is the dual Coxeter number.
 If the $(k+h^V) \neq 0$ we define the Virasoro operators
$L_n$
  by normalizing
  $l_n$ as follows  $L_n \equiv (k+h^V)^{-1} l_n$. The $L_n$s obeys the
standard
   Virasoro algebra
 \beq
  [L_n, L_m] = (n-m) L_{n+m} + \frac{k \dim{\bf g}}{12 (k +
h^V)}(n^3-n)\delta_{n+m}.
 \eeq{virasoro}
  As a result, the world-sheet (properly normalized) Hamiltonian has the
form
  \beq
   L_0 = \frac{1}{2(k+h^V)}\sum\limits_{m = -\infty}^{+\infty} : J^A_m
 \eta_{AB} J^B_{-m}:.
  \eeq{Lo}
 Thus we conclude that the quantum tension is $(k+h^V)$ rather than
$k$.
  Note that unitarity (i.e., the positivity of the central charge) puts a bound on
 the level, $k$: $-h^V < k < \infty$
  and that we normalize our objects such that $k$ is positive.

    For example, for $SL(2,R)$ the unitarity bound is:
     $2 < k < \infty$. Therefore
      $k$ cannot be taken to zero in the quantum
theory. However, we suggest that the limit $k \rightarrow 2 $
represents the tensionless limits in the quantum theory.
 The central charge of the model is
\beq
 c = \frac{3k}{k-2}
\eeq{centrcha}
 and thus to embed the model into the critical bosonic string theory the following
 bound should be satisfied: $c \leq 26$ (i.e., $k \geq 13/2$). Therefore one may think that
 the limit  $k \rightarrow 2 $ cannot be done within the critical string theory.
 However it may happen that in the neighborhood of the point $k=-h^V$ the theory should be
 redefined and the Virasoro algebra is not relevant anymore. Having this in mind
 we will ignore this problem in what follows.

  When the level $k$ is equal to $-h^V$ (i.e., for $SL(2,R)$ when $k=2$) it
is called the
   critical level. In mathematics WZW models at critical level
have attracted
    a lot of attention and the representation theory of corresponding
affine
     algebra has been considered in  \cite{Feigin:1991wy}. However in the present
context we are
      interested in the possibility to interpret a noncompact WZW model
at critical level
       as an (unconventional) string theory, possibly  with a big new
symmetry.

   Let us state some relevant properties of WZW models at
 critical level. At the critical level we cannot introduce $L_n$s which
obey the Virasoro algebra. However, there are still Sugawara
operators $l_n$
 which commute with each other and with $J^A_n$
 \beq
  [l_n, J_m^A] =0,\,\,\,\,\,\,\,\,\,\,\,\,\,\,\, [l_n, l_m]=0 .
 \eeq{critcomrel}
 Thus, at the critical level there is a large number of new null states ,
e.g.,
 all states of the form
  \beq
   l_{n_1} l_{n_2} ... l_{n_p} |0,\alpha\rangle,\,\,\,\,\,\,\,n_i >
 0,\,\,i=1,...,p
  \eeq{nulstrcurvm}
  are null states. In (\ref{nulstrcurvm})  $|0,\alpha\rangle$ is a
state
   with the property that $J^A_n|0,\alpha\rangle =0$ for $n > 0$ and
$\alpha$ is
    a label for a finite dimensional representation of ${\bf g}$, $J_0^A|0,\alpha\rangle =
 \alpha^A |0,\alpha\rangle$.
     Compared to the noncritical level, the number of zero-norm states
increases dramatically
    when $k=-h^V$  thus indicating the appearance
     of the gauge symmetry of the space-time theory we are seeking.

 Another important question is what would happen with the Virasoro
constraints
  in this limit (i.e., $k \rightarrow - h^V$).
   Naively  the constraints will collapse to the following ones
  \beq
   l_n |phys \rangle = 0,\,\,\,n \geq 0 .
  \eeq{constgrop}
 Using the properties (\ref{critcomrel}) the conditions
(\ref{constgrop})
  becomes just the single condition
  \beq
   C_2 |phys \rangle = 0 ,
  \eeq{Casimir}
 where $C_2$ is a quadratic Casimir.
  However,this reasoning may be too naive. Let us look at the subset
  consisting of
 the following states
\beq
 |\epsilon\rangle = \epsilon_{A_1 A_2 ...A_n} J_{-1}^{A_1} J_{-1}^{A_2} ... J_{-1}^{A_n}
 |0,\alpha\rangle
\eeq{newstates}
 where $\epsilon_{A_1 A_2 ...A_n}$ is completely symmetric tensor. Away from critical level
 the Virasoro conditions imply that
\beq
 C_2(\alpha)= (k+h^V) (n-1),\,\,\,\,\,\,\,\,\,\,\,\,\,\, \epsilon_{A_1 A_2 ...A_n} \alpha^{A_1} =0,
\eeq{Vircondimpl1}
 where $C_2(\alpha)$ is a quadratic Casimir for the representation $\alpha$. 
 When the level $k$ goes to the critical value $-h^V$ the states (\ref{newstates}) become massless (i.e.,
 $C_2(\alpha)=0$), but the transversality condition remains true. Thus at the critical level we reproduce the
 analogue of the Fronsdal's conditions in the fixed gauge. However at the critical level the
 trasversality condition does not arise from $l_n$, which it does in the flat case.

 We now formalize the tensionless limit may  somewhat.
 In flat space we  scaled  the zero and non-zero modes differently with respect to
$\alpha'$. Thus the flat space tensionless limit can be formulated
as follows:
 We introduce a  parameter $R$ and rescale the parameters of the theory as
 $\eta_{\mu\nu} = R^{-1}\tilde{\eta}_{\mu\nu}$, $a_n^\mu = \sqrt{R} \tilde{a}_n^\mu$ and
 $p_\mu = \tilde{p}_\mu$ (i.e., we do not scale the {\em contravariant} zero mode).
 The limit $R \rightarrow \infty$ gives rise to the tensionless limit and it
 does not change the underlying Heisenberg algebra. However the Virasoro algebra ($L_0 = Rl_0$
 and $L_n = \sqrt{R} l_n$, $n\neq 0$) gets contracted to the algebra (\ref{HEalg}).

 Let us now turn to the affine algebra (\ref{KMalg}) and try to apply the same
  logic\footnote{We thank Ergin Sezgin for a valuable discussion of
 this issue.}.
  We have to  rescale the zero and the nonzero modes as well as the
   metric $\eta_{AB}$ in some way ( and, as a result, we also have to scale the
 structure constants since $f^{AD}_{\,\,\,\,\,\,\,\,C}f^{BC}_{\,\,\,\,\,\,\,\,D}\sim
 \eta^{AB}$). There is a scaling which would preserve the affine
 algebra: $\eta_{AB} = R^{-1}\tilde{\eta}_{AB}$, $f^{AD}_{\,\,\,\,\,\,\,\,C}
 = \sqrt{R} \tilde{f}^{AD}_{\,\,\,\,\,\,\,\,C}$, $J^A_n = \sqrt{R} \tilde{J}^A_n$ and
 $J_{0}^{A} = \sqrt{R}\tilde{J}_{0}^{A}$. However, this scaling does nothing with the Virasoro generators
 and it does not lead to anything new.

Next we can try to mimic the flat space case by scaling the zero
and non-zero modes differently.
 In particular we can keep fixed the contravariant zero mode $J_{0A}=\tilde{J}_{0A}$ ($J_{0}^A= R\tilde{J}_{0}^A$)
 and scale the rest as before. We then obtain the
  following conditions on the physical states in the limit $R \rightarrow \infty$
  \beq
   ( \tilde{J}_0^A \tilde{\eta}_{AB} \tilde{J}^B_n) |phys\rangle = 0,\,\,\,\,n \geq 0.
  \eeq{newVirgr}
 However the algebra (\ref{KMalg}) does not have a well defined limit in this case.
  We may continue and study other scalings  of the affine algebra with well-defined limit.
 Typically the limit will lead to the contraction of the affine algebra and thus it will
 change the model drastically.

 Although we were able to reproduce the Fronsdal's conditions at the critical level we could
 not derive the ``Virasoro'' conditions responsible for the transversality condition (\ref{Vircondimpl1}).
Indeed it is tempting to treat $(k+h^V)^{-1}$ as we treated
$\alpha'$ in the flat case. However for the group manifold we
cannot perform the limit as in the flat space
 example considered in previous section.
More
     physical intuition is needed to find the right
prescription
      for the limit. In the next section we consider a
more physical
       example, where we know what to expect from the
tensionless
        string spectrum.

 \section{Tensionless strings in $AdS_d$}
 \label{s:coset}

 In this section we consider string theory over coset manifolds
  based on noncompact groups. The logic to a large extent follows that in the previous
   section. First we discuss
    $AdS_d$ as a specific example.

 Following the work of Fradkin-Linetsky,  \cite{Fradkin:mw}, \cite{Fradkin:ie} we represent $AdS$-space
  as a coset symmetric space of the form
 \beq
    AdS_d = \frac{SO(d-1,2)}{SO(d-1,1)},
 \eeq{defADS}
  where $SO(d-1,2)$ is the anti-de Sitter group in d dimensions and $SO(d-1,1)$ is its Lorentz
   subgroup. The underlying CFT may be thought of  as a $SO(d-1,2)$ WZW model with
    gauged subgroup $SO(d-1,1)$.
    We ignore the questions of consistency of this
     theory and the fact that there are other proposals
      for the string theory in $AdS_d$, \cite{deBoer:1999ie}.
     Our intension is to give a rough idea of how things may work,
      and this does not rely on the particularities  of the coset constructions.

  The
 affine Kac-Moody algebra $SO(d-1,2)$ of the $AdS_d$ coset model is of the form
 \beq
 [M_n^{\mu\nu}, M_m^{\rho\sigma}]= i(\eta^{\mu[\sigma|}
 M_{n+m}^{\nu|\rho]}
   + \eta^{\nu[\rho|} M_{n+m}^{\mu|\sigma]}) - k n
 (\eta^{\mu\rho}\eta^{\nu\sigma}
     - \eta^{\mu\sigma} \eta^{\nu\rho})\delta_{n+m}
 \eeq{MMcomm}
 \beq
 [M_n^{\mu\nu}, P^\rho_m] = i (\eta^{\nu\rho} P^\mu_{n+m}
 -\eta^{\mu\rho} P^\nu_{n+m})
 \eeq{PMcomm}
 \beq
 [P_n^\mu, P_m^\nu] = iM^{\mu\nu}_{n+m} + k n \eta^{\mu\nu}
 \delta_{n+m}
 \eeq{PPcom}
 where $k = T \Lambda^{-1}$ and $(-\Lambda)$ is the  cosmological constant in $AdS_d$.

 The Virasoro generators are constructed according to standard Goddard-Kent-Olive
  construction
 \beq
  L_n = \frac{1}{2} \sum\limits_{m=-\infty}^{\infty}
   \left [ \frac{1}{k-(d-1)} :P_m^\mu \eta_{\mu\nu} P_{n-m}^\nu:
   + \left(\frac{1}{k-(d-1)} - \frac{1}{k-(d-2)}\right ) :M_m^{\mu\nu} \eta_{\mu\rho}
   \eta_{\nu\sigma} M_{n-m}^{\rho\sigma}:\right ]
 \eeq{stress}
 where we used the fact that for $SO(d-1,2)$ and $SO(d-1,1)$ the dual Coxeter numbers
  are $(d-1)$ and $(d-2)$ respectively. Thus the level is bounded by
   the values: $(d-1) < k < \infty$. The limit $k \rightarrow \infty$
    corresponds to the flat space limit (with the affine currents
     appropriately rescaled). The tensionless limit would correspond to
      $k \rightarrow (d-1)$, and as before we get a dramatic increase in the number
       of zero-norm states (they will be constructed out of Sugawara tensors
       for $SO(d-1,2)$ with positive $n$).

For generic noncritical $k > (d-1)$ the theory has $SO(d-1,1)$ global symmetry
 since
\beq
 [L_0, M_0^{\mu\nu}] = 0
\eeq{noncritsym}
 where $L_0$ is the worldsheet Hamiltonian. Thus all states of the theory are organized in 
 the representations of $SO(d-1,1)$. If, in analogy with the flat limit, we define 
 the tensionless limits as the massless limit of the theory (i.e., the limit when all 
 states become massless) then we should have an enlargement of the symmetry to 
 the AdS-group, $SO(d-1,2)$. However this never happens at noncitical values of $k$. 
 Thus if the tensionless limit exists it must correspond to the theory at the critical 
 level $k=(d-1)$ where 
 \beq
 [l_0, M_0^{\mu\nu}]=0,\,\,\,\,\,\,\,\,\,\,\,
 [l_0, P_0^\mu] = 0 .
\eeq{critsym}
 $l_0$ is the worldsheet Hamiltonian at the critical level, related
 to the noncritical as follows $(k-(d-1))L_0 = l_0$. From this simple argument 
 we conclude that if the tensionless limit exists then it should be at the critical 
 level limit. At the critical level we will thus necessary have higher spin 
 massless states, unless the theory becomes trivial. To investigate if the tensionless
 theory is trivial or nontrivial one should study the behavior of the string 
 spectrum in the vicinity of the critical level. This problem seems to be 
 hard. Indeed nothing is known about the Regge trajectories in this coset model. 
 We hope to come back to this question in the future.

There is another important point to be addressed. The tensionless limit of string 
 theory and the appearance of massless higher spin states in the string theory are not 
 equivalent notions. The tensionless limit (if it exists) implies the existence of massless higher 
 spin states. However the presence of higher spin states does mean that the theory is
 in tensionless phase. The possibility that  massive higher spin states 
  coexist with massless cannot be excluded. 
 For example, in the present model of $AdS_d$ we know that there are no higher spin states
 in the semiclassical regime (i.e., when $k$ is big), but we have very little knowledge of what happens when $k$  moves towards the critical value. In particular, we do not know if 
massless higher spin states  appear at some value of $k$. However if they arise 
 at $k \neq (d-1)$ they will mix with the massive states. The completely 
massless spectrum will appear only at $k=(d-1)$. 
     
We finally note, that in units where $\alpha ' =1$, the critical level may be interpreted as
a critical radius of $AdS_d$. According to our discussion the symmetry at this radius is greatly enhanced.

 \section{Summary and discussion}
 \label{s:end}

 In this short note we have discussed the tensionless limit in the quantum
 string theory at the level of the Fock space and the Virasoro constraints on physical states.
We found that in flat space the truncated Virasoro constraints correctly reproduce
  the Fronsdal's conditions for free higher spin massless fields.
  We then applied the same type of procedure in curved manifolds, in particular
  in $AdS_d$ space.

 In curved space we found that there is a critical value of string tension
  related to the critical value of the level of WZW model. The theory has
   very special properties at the critical tension where the  number of null states
    drastically increase, indicating the appearance of
    a  very large gauge symmetry in the underlying space-time theory.

     At present it is not clear if there is a similar critical tension for
      superstrings.\footnote{Superstrings do seem to have an important relation
to higher spin theories, however, see, e.g.,
\cite{Sezgin:2002rt} } Presumably this depends on the properties of the background under consideration.

Another interesting question is
        the $AdS/CFT$ interpretation of the critical tension (assuming it exists
         for $AdS_5\times S^5$). The expression $g_{YM}^2 N = (R\sqrt{T})^4$
          relates the Yang-Mills coupling to the radius of $AdS_5$ (and $S^5$).
       The existence of critical value for $g^2_{YM}N$ seems unlikely. Therefore
        {\em if} there is a critical tension then, in analogy to $k\to k+h^V$, it should lead to a modification
         of the expression according to  $g_{YM}^2 N = (R\sqrt{T} + (R\sqrt{T})_{crit})^4$.
  Previously the existence of critical string tension has been argued and its relation to higher
 spin theories discussed in \cite{Sezgin:2002rt}. Further supportive argument in favour of critical tension
 has been recently considered in \cite{Bianchi:2003wx}.

The obvious future directions of this investigation are  to extend the discussion
 to superstrings as well as to make more rigorous some of the qualitative arguments presented here.

 \bigskip

 \bigskip

 {\bf Acknowledgements}: We are grateful to Massimo Bianchi, J\"urgen Fuchs, Joseph Minahan,
 Augusto Sagnotti,
 Ergin Sezgin, Konstadinos Sfetsos, Per
 Sundell,  Mikhail Vasiliev and Konstantin Zarembo
   for discussions.
   MZ would like to thank the Institute of Theoretical Physics,
Uppsala
 University,
    where this work was carried out.
 UL acknowledges support in part by EU contract
 HPNR-CT-2000-0122 and by VR grant 650-1998368.
 MZ acknowledges support in part by EU contract
    HPRN-CT-2002-00325.

 \end{document}